# Controlled phase separation in La$_{0.5}$Ca$_{0.5}$MnO$_3$

P. Levy,* F. Parisi, G. Polla, D. Vega, G. Leyva, and H. Lanza
*Departamento de Física, Comisión Nacional de Energía Atómica, Gral Paz 1499 (1650) San Martín, Buenos Aires, Argentina*

R. S. Freitas and L. Ghivelder
*Instituto de Física, Universidade Federal do Rio de Janeiro, 68528, Rio de Janeiro, RJ21945-970, Brazil*

A systematic study of phase separation effects in polycrystalline La$_{0.5}$Ca$_{0.5}$MnO$_3$ obtained under different thermal treatments is reported. Samples with average grain size ranging from 200 to 1300 nm were studied. Magnetic and electrical measurements show quantitative differences among samples in their low-temperature behavior, indicating that the fraction of the ferromagnetic (FM) phase gradually decreases as the grain size increases. Percolation of the FM phase in samples with a small fraction of this phase suggests that grain boundaries play a distinctive role in the spatial distribution of coexisting phases. The defective structure at the grain surface could explain the local inhibition of the antiferromagnetic charge ordered phase, an effect that is gradually removed with increasing grain size. Qualitative agreement of the data with this description is found. This effect is also found to be highly dependent on the oxygen content of the samples and its spatial distribution.

## I. INTRODUCTION

The hole-doped rare-earth manganites $L_{1-x}A_x$MnO$_3$ ($L$ is a lanthanide and $A$ a divalent alkaline earth) are being the focus of extensive investigation. The strong interplay between electronic, magnetic and structural properties displayed by these materials gives rise to a wide variety of phases.[1] Most of the interesting physical properties of the manganese perovskites arise from the competition between ferromagnetic double exchange and antiferromagnetic superexchange, the ratio of these competing interactions being determined by intrinsic parameters such as doping level, average cation size, cation disorder and oxygen stoichiometry.

The system La$_{1-x}$Ca$_x$MnO$_3$ has a rich phase diagram,[2] where paramagnetism ($P$) ferromagnetism (FM), antiferromagnetism (AFM), and orbital and charge ordering (CO) are determined by the temperature and the doping level $x$. Its ground state is ferromagnetic metallic (FMM) for $0.15 < x < 0.5$. The phase boundary point $x = 0.5$ is the focus of great interest. Following early reports, upon lowering temperature this compound first undergoes a $P$ to FM phase transition at $T_C \simeq 225$ K, and then to a COAFM phase at $T_{CO} \simeq 155$ K.[3,4] Nevertheless, experimental data on magnetization and resistivity do not always agree with that description and slight differences from one to other report can be found in the literature. These disagreements are found mainly in the low temperature region, where a residual magnetization of some tenths of $\mu_B$/Mn is observed,[2,3,5,6] and metallic like behavior below $T_{CO}$ is sometimes obtained.[7,8] The fact that, in general, a nonfully AFM state is reached at low temperatures has been early attributed to small variations in cation or oxygen stoichiometry, or to the existence of a canted antiferromagnetic phase, but recent results from nuclear magnetic resonance data,[8–10] electron microscopy,[11] neutron scattering,[4] and magnetostriction[7] show a more complex scenario, in which the low-temperature state of La$_{0.5}$Ca$_{0.5}$MnO$_3$ is characterized by the coexistence of FMM and COAFM phases at the microscopic level. This residual low-temperature magnetization would then be a measure of the fraction of the total volume that corresponds to the FM phase which is trapped in the AFM host, and the metallic behavior a consequence of the formation of percolative paths of FMM clusters across the sample.

The origin of the inhomogeneities in the magnetic properties is not clear. Recently, Uehara et al.[12] presented a systematic study of the coexistence of phases in La$_{5/8-z}$Pr$_z$Ca$_{3/8}$MnO$_3$; they showed that the relative fraction of the coexisting phases can be controlled by the Pr content $z$, and explained the metallic behavior by percolation of FM submicrometer-scale domains. In this system, competition between the FM ground state of the $z = 0$ compound and the COAFM ground state of the $z = 5/8$ one seem to be the reason of the observed features.

In La$_{0.5}$Ca$_{0.5}$MnO$_3$, the stable state between competing FM and COAFM phases is determined by the temperature. Although the COAFM phase is the ground state, its small energy difference with the FM state, revealed by magnetic[6] and time relaxation measurements[13] can indicate a tendency towards phase coexistence. There are some trends in the literature pointing to an explanation of this feature in the framework of the electronic phase separation scenario predicted for the manganites[14] but it is more probable that such behavior can be mainly determined by structural inhomogeneities characteristic of the ceramic samples, which can enhance or even determine local magnetic properties. In this picture it appears that average quantities, as mean cation size, bond angles, etc., could not be the appropriate parameters to account for physical properties which can be determined on a short length scale by, for instance, cation disorder[15] and off-stoichiometric oxygen distribution.[16] Neutron scattering experiments[4] have shown that the COAFM state can be described as two interpenetrating lattices of Mn, each one with a different coherence length (a few hundred Å for the Mn$^{+3}$ sublattice, some thousands Å for the Mn$^{+4}$ one). Then, physical properties could be dramatically changed by structural inhomogeneities at that scale.

TABLE I. Thermal treatments performed on samples discussed in the text, their mean grain size and percentage of $Mn^{+4}$ content (absolute error 2%).

| Sample | Thermal treatment | Grain size (nm) | $Mn^{+4}$ (%) |
|---|---|---|---|
| A  | 5 h at 900 °C       | 180  | 54.5 |
| B  | A+5 h at 950 °C     | 250  | 53.4 |
| C  | B+5 h at 1000 °C    | 450  | 55.6 |
| D  | C+5 h at 1100 °C    | 950  | 53.5 |
| E  | D+5 h at 1200 °C    | 1300 | 55.3 |
| E1 | E+5 h at 1200 °C    | 1300 | 54.0 |
| E2 | 10 h at 1200 °C     | 1300 | 51.4 |

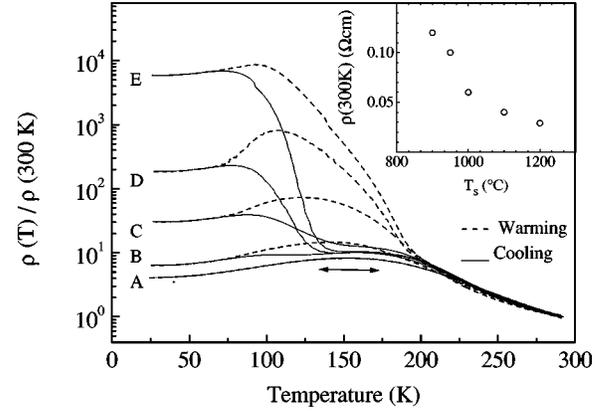

FIG. 1. Normalized resistivity as a function of temperature for samples $A$–$E$. The inset displays room temperature resistivity $\rho(300\,K)$ as function of the final sintering temperature $T_s$.

Grain boundaries appear as the most drastic disrupt of an idealized perfect crystal, owing to their inherent character, but also because they can act as accumulative pinning centers for structural defects. The dependence of the magnetic properties on grain size has been established in several related compounds,[17,18,19,20,21] and the differential behavior of surfaces respect to bulk material has been determined.[22] These studies were performed on systems with a well-defined ground state (FM); the role of grain boundaries is expected to be drastically enhanced in compounds with a mixed phases character. In this work we study the magnetic and transport properties in submicronic samples of $La_{0.5}Ca_{0.5}MnO_3$. We found that low-temperature resistivity and magnetization are strongly dependent on grain size, but also on other sources of defects, such as oxygen content and its spatial distribution. We present evidence that using a specific thermal treatment process it is possible to control the FM-COAFM coexistence in a nearly continuous way, leading to a wide range of values for the low temperature resistivity and magnetization.

## II. EXPERIMENTAL

Polycrystalline samples were obtained by a citrate/nitrate decomposition method using 99.9% purity reactants. After the mixed citrates were dried at 100 °C they were heated in air at 700 °C for 15 h, and then furnace cooled to room temperature. Additional thermal treatments were performed in air. We have prepared two sets of samples. On one hand, samples with different average grain size were obtained by performing short time (5 h) accumulative thermal treatments on the same batch with a gradual increase of the final sintering temperature. As magnetic and transport properties of $La_{1-x}Ca_xMnO_3$ are critically dependent on both the doping level $x$ and the oxygen content in the region near $x=0.5$, we have performed this particular thermal process with the twofold aim of assuring the same $x$ for all samples and minimizing changes in the oxygen content as the grain size is increased. Samples obtained in this way are labeled $A$ to $E$. On the other hand, two additional samples ($E1$ and $E2$) were obtained through different thermal processes, originating samples with different thermal history but the same average grain size. Thermal history of different samples is presented in Table I. In all cases samples were cooled down to room temperature at about 100 °C/h. The phase purity, unit cell dimensions and structural parameters were analyzed using the Rietveld method. XRD data were gathered at room temperature in a Bragg-Brentano diffractometer using CuK$\alpha$ radiation, 0.5° dispersion slit, step size 0.02° and counting time 25 sec by step. Four probe resistivity measurements were performed in the temperature range 30–300 K on polycrystalline pellets previously pressed and sintered. $Mn^{+4}$ contents were determined by iodometric titration. Magnetization measurements were performed in a commercial magnetometer (Quantum Design PPMS) between room temperature and 2 K with applied fields up to $H=9$ T. Average grain size was estimated through SEM microphotographs.

## III. RESULTS AND DISCUSSION

X-ray diffraction patterns of as grown and thermal treated samples showed them to be single phase and all the patterns could be indexed on the basis of an orthorhombic cell with space group $Pnma$. No changes in peak positions were observed through the series, indicating that cell parameters are nearly the same for all samples [$a=5.4148(5)$ Å, $b=7.6389(7)$ Å, $c=5.4260(5)$ Å]. Broadened reflections (0.6°) were found due to small particle size in sample $A$ with processing conditions performed at 900 °C. On the other hand, narrow peaks (0.2°) were found for sample $E$, with a final heating temperature of 1200 °C. The percentage of $Mn^{+4}$ was found almost constant through the series (around 54%).

Figure 1 shows the normalized resistivities $\rho$ for samples $A$ to $E$ as a function of temperature. The resistivity of sample $A$ displays typical features of a $P$- FM reversible system, with activated conduction at high temperatures, and metalliclike behavior at low temperatures, without any signature of charge order. The resistivity of sample $B$ follows that of sample $A$ down to $T=125$ K on cooling, where an increase is observed, indicating the presence of charge order. Near $T_p=100$ K a new peak develops and metallic behavior is obtained at lower temperature. Irreversibility between 70 and 180 K is revealed on warming. Overall features of sample $B$ are observed in samples $C$, $D$, and $E$, with a gradual increase of the low temperature resistivity values and a decrease of $T_p$.

Magnetization ($M$) data measured as a function of tem-perature with $H=1$ T on samples $A$ to $E$ are shown in Fig. 2. All of them show a FM ordering at the same temperature $T_C \simeq 250$ K but, while sample $A$ apparently remains FM in all the temperature range, a clear FM-AFM transition is observed for the other samples. The low-temperature magnetization changes smoothly from $3\mu_B$/Mn for sample $A$ to $0.3\mu_B$/Mn for sample $E$, showing that a FM phase is coexisting with AFM in all samples, but in different proportions.

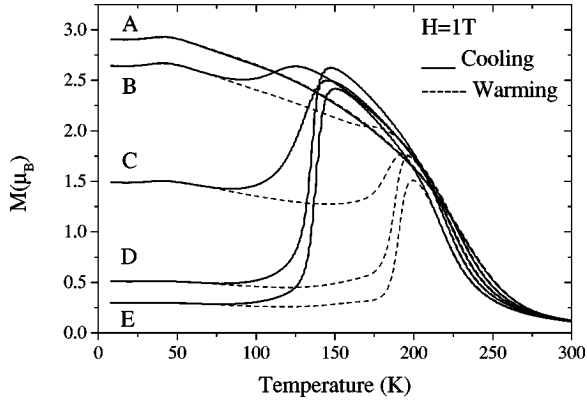

FIG. 2. Temperature dependence of the magnetization ($H=1$ T) for samples $A-E$.

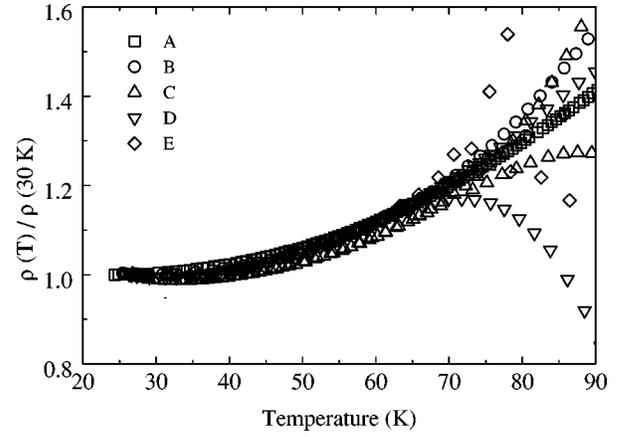

FIG. 4. The temperature dependence of the resistivity for samples $A-E$ normalized by their value at 30 K.

Measurements of $M$ vs $H$ at 10 K allow an estimation of the percentage of FM phase in the samples. The behavior due to each phase is clearly separated in a graph of $M$ vs $1/H$, plotted in Fig. 3. This field dependence arises from the dominant term in the "law of approach to saturation."[23] The straight lines in the graph are extrapolated to high fields ($1/H \to 0$) to give the FM saturation moment, which can be compared with the theoretical value for a fully FM sample ($3.5\mu_B$/Mn). We obtain 84, 77, 54, 15, and 9 for the percentage of the FM phase in samples $A$ to $E$, respectively. The upturn at small values of $1/H$ signals the presence of the AFM phase. This feature is present even in sample $A$, which has no signal of the presence of COAFM phase in either $\rho(T)$ or $M(T)$ data.

In spite of quantitative differences in the low temperature behavior, the Curie temperature $T_C$ and $T_{CO}$ are nearly the same throughout the series, and the samples display the same ferromagnetic behavior around $T_C$ (Fig. 2). All samples display a common metallic behavior in the low-temperature range: their resistivity curves for $T<70$ K collapse into a single one when normalized by their resistivity at 30 K, $\rho(30$ K$)$, as shown in Fig. 4. This behavior can be doubtless related with the percolation of a single FM phase. Quantitative changes in the resistivity values for the different samples are then due to changes in the fraction of the material that remains FM and percolates. In the simplest model each sample can be considered as a parallel circuit formed by the fraction $y$ of the total volume which is FM and percolates (characterized by resistivity $\rho_F$) and $1-y$ of the rest of the material (characterized by $\rho_{CO}$) consisting of COAFM and nonpercolating FM phases. The total resistivity $\rho$ can then be written as

$$1/\rho = y/\rho_F + (1-y)/\rho_{CO}.$$

With this picture it is easy to show that, in the temperature range where $\rho_F/\rho_{CO} \ll y$ the function $\rho/\rho(30$ K$)$ is $y$-independent (i.e., sample independent), in agreement with experimental results shown in Fig. 4. Of course, the validity of this argument also requires $y$ to be almost temperature independent in that interval, as expected in the low-temperature reversible region. It is worth noting that a similar model was used by Roy et al.,[6] who studied the metallic behavior induced by application of moderate magnetic fields in an otherwise insulating sample of La$_{0.5}$Ca$_{0.5}$MnO$_3$. They found the same common behavior for the normalized low-temperature resistivity for different applied fields. Their description, in terms of a field dependent number of free carriers $n(H)$, is obviously related to the percolating fraction $y$ of the FM phase. Surprisingly, they found no percolation for $H=0$, although the low-temperature magnetization was around $1\mu_B$/Mn for $H=1$ T. This indicates an isolated cluster distribution for the FM phase, as that observed in Ref. 11. The application of a magnetic field increases the volume of the clusters, and eventually leads to the percolation of the metallic phase.

In our case the change in the fraction $y$ of the percolative FM phase is controlled through the thermal treatment pro-cess. Assuming $y=1$ for the almost pure FM sample $A$, we obtain $y \approx 0.64, 0.4, 0.05$, and $0.002$ for samples $B$, $C$, $D$, and $E$ respectively. As the magnetization of sample E at low temperatures is about 10% of the saturation value of sample A, only a 2% of this minority FM phase percolates (i.e., 0.2% of the total sample volume). As a comparison, in Ref. 6 percolation is obtained through 1% of the sample by applying $H=9$ T. All these facts imply that in phase separation systems the relationship between low-temperature magnetization and transport properties is not direct, and not only the presence but also the spatial distribution of the metallic phase must be considered.

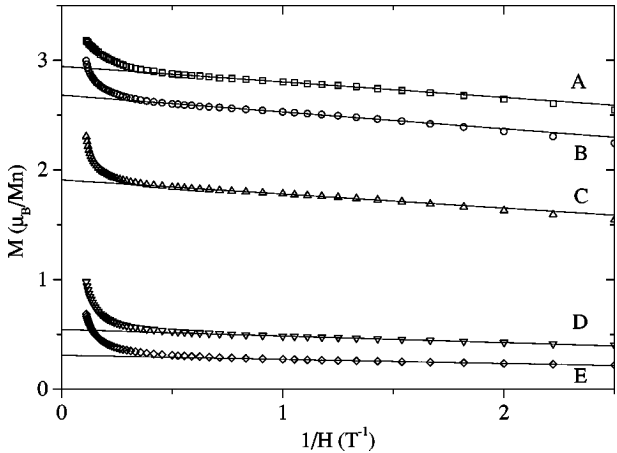

FIG. 3. Magnetization as a function of $1/H$ at 10 K for samples $A-E$. The extrapolation of the straight lines indicates the saturation values of the FM phase. The fraction of the FM residual phase was calculated assuming a maximum value of $3.5\mu_B$/Mn.

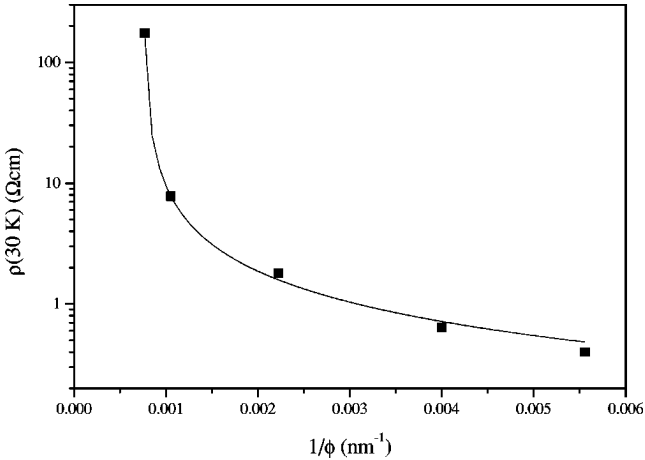

FIG. 5. $\rho(30\text{ K})$ data for samples $A-E$ as a function of the inverse grain size. The solid line is the adjustment of the data points with the function $\rho(30\text{ K}) = A(1/\phi - 1/\phi_c)^{-1}$, where $A = 0.0023\,\Omega$ and $\phi_c = 1323$ nm.

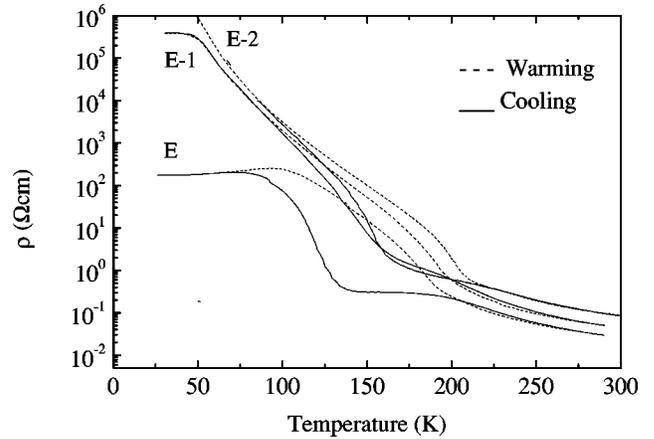

FIG. 6. Normalized resistivity as a function of temperature for samples $E$, $E1$, and $E2$ on cooling and warming.

In the $La_{5/8-z}Pr_zCa_{3/8}MnO_3$ system the percolation of the FM phase is explained by a phase transition which changes the system from a state of coexisting short ranged CO and FM nanodomains to a phase characterized by the coexistence of long range FM with CO domains.[12] Instead, in $La_{0.5}Ca_{0.5}MnO_3$ no additional phase transition related to percolation is observed: transport properties are determined by the spatial distribution of the untransformed high-temperature FM phase trapped in the COAFM host below $T_{CO}$. Percolation of the metallic phase even in those samples with a small fraction of FM material point to the grain boundaries as a candidate to produce the aggregation of the FM phase. At grain boundaries strain and local compositional variations characteristic of surfaces can give rise to a zone in which the structural disorder causes the local inhibition of the COAFM phase.

Within this simplified picture, each grain can be described by an insulating core and a FMM surface layer of thickness $\delta$, so a $\delta/\phi$ dependence for the percolative fraction $y$ with the grain size $\phi$ is expected. In Fig. 5 we sketch $\rho(30\text{ K})$ vs $1/\phi$. Experimental data can be well fitted by the function $\rho(30\text{ K}) = A(1/\phi - 1/\phi_c)^{-1}$. This functional form implies $y \sim (1/\phi - 1/\phi_c)$; thus a $\delta/\phi$ dependence for $y$ is obtained if the thickness $\delta$ of the FMM surface layer diminishes linearly with increasing grain size. The parameter $\phi_c \sim 1320$ nm can be then interpreted as the critical grain size for which the volume effect suppresses surface disorder.

The weakening of the surface layer can be due to both intrinsic and extrinsic grain size effects. The former was shown to have a strong influence on magnetic properties and magnetoresistance response of manganites;[18] in particular, this effect has been described by the decrease of the thickness of the disordered layer as grain size increases in $La_{2/3}Sr_{1/3}MnO_3$.[20] On the other hand extrinsic effects as oxygen content,[6] nonuniform oxygen distribution[24] or surface contamination[25] were recently shown to be relevant parameters to describe transport properties on related compounds.

We have indications that the oxygen content in the samples does not change: Macroscopic parameters related with it such as $T_C$ and cell constants do not show substantial changes with thermal treatments, and the $Mn^{+4}$ measurements gave values close to 54% for all samples. As a deoxygenation process involves cation diffusion it is feasible that our short-time accumulative annealing step (5 h) is not enough to produce substantial changes in the oxygen stoichiometry. However, the trend of the compound to loss the oxygen excess as the sintering temperature is increased can lead to an internal diffusion of vacancies towards the surface, yielding a nonuniform distribution of the off-stoichiometric oxygen. This effect can be enhanced by the accumulative thermal treatment process performed here, which yields the increase of the grain size by fusion of the grains obtained in the previous step. Thus, new defects at the glued boundaries are generated which, in turn, will be affected by the diffusion process.

In order to clarify this point, we have performed magnetization and resistivity measurements on a new series of samples, all of them with the same average grain size (1300 nm) but obtained through different thermal treatments. Sample $E1$ was obtained by an additional 5 h thermal treatment at 1200 °C of sample $E$, and sample $E2$ by a 10 h thermal treatment at 1200 °C of the as-grown powder with no intermediate steps. Figure 6 shows the resistivity data for samples $E$, $E1$, and $E2$. A degraded metallic behavior is displayed by sample $E1$ with respect to sample $E$. This may indicate that, keeping constant the grain size, the additional heating produces the cleaning of the surface through elimination of oxygen excess, weakening the thickness of the FMM layer. On the other hand, no metallicity is observed in the single-step grown sample $E2$, which displays the insulating behavior of a pure COAFM system at low temperature.

The low temperature magnetization of sample $E2$ at 1 T is also around $0.3\mu_B$/Mn, thus confirming the presence of a FM phase not distributed along percolative paths. These results show that the grain size dependence of the physical properties is highly influenced by other factors, as the actual oxygen stoichiometry or spatial defect distribution. Thus, series with different oxygen content would be described by different values of the critical grain size parameter $\phi_c$.

## IV. CONCLUDING REMARKS

The overall presented data confirm that the coexistence of FM and COAFM phases is affected by the local defect structure. In the submicrometer samples studied here the low-temperature metallic behavior is obtained because the COAFM state is partially inhibited by structural disorder at the grain surfaces, which acts as an accumulation point for these defects. Changes in low-temperature resistivity values as thermal treatments are accumulated can be ascribed to both grain size and oxygen distribution effects. Beyond the detailed mechanisms determining phase separation, the FMM-COAFM coexistence can be controlled in an almost continuous way giving rise to a wide range of possible values for the low-temperature resistivity and magnetization. Phase separation has been recently signalized as the responsible for the large low-temperature magnetoresistance effect observed in $La_{5/8-z}Pr_zCa_{3/8}MnO_3$. The possibility of controlling the coexistence of phases in $La_{0.5}Ca_{0.5}MnO_3$ opens a route to study the phase separation effects on magnetoresistance without changing $T_C$ and $T_{CO}$; this could give additional insight about the mentioned interrelation.

## ACKNOWLEDGMENTS

We thank L. Civale for help during preliminary magnetization measurements. Part of this work was supported by CONICET PEI 98 No.125.